\documentclass[twocolumn,showpacs,preprintnumbers,amsmath,amssymb,pra]{revtex4-1}
\usepackage{graphicx}% Include figure files
\usepackage{dcolumn}% Align table columns on decimal point
\usepackage{bm}% bold math
\usepackage{color}
\begin{document}

\title{On-Demand Generation of Traveling Cat States Using a Parametric Oscillator}

\author{Hayato Goto,$^1$ Zhirong Lin,$^2$ Tsuyoshi Yamamoto,$^{2,3}$ and Yasunobu Nakamura$^{2,4}$}
\affiliation{
$^1$Frontier Research Laboratory, 
Corporate Research \& Development Center, 
Toshiba Corporation, 
1, Komukai Toshiba-cho, Saiwai-ku, Kawasaki-shi, 212-8582, Japan
\\
$^2$RIKEN Center for Emergent Matter Science (CEMS), Wako, Saitama 351-0198, Japan
\\
$^3$System Platform Research Laboratories, NEC Corporation, Tsukuba, Ibaraki 305-8501, Japan
\\
$^4$Research Center for Advanced Science and Technology (RCAST), The University of Tokyo, Meguro-ku, Tokyo 153-8904, Japan}

\date{\today}

\begin{abstract}

We theoretically propose a method 
for on-demand generation of traveling Schr\"odinger cat states,
namely, quantum superpositions of distinct coherent states of traveling fields.
This method is based on deterministic generation of intracavity cat states 
using a Kerr-nonlinear parametric oscillator (KPO)
via quantum adiabatic evolution.
We show that the cat states generated inside a KPO 
can be released into an output mode by controlling the parametric pump amplitude dynamically.
We further show that the quality of the traveling cat states can be improved
by using a shortcut-to-adiabaticity technique.

\end{abstract}

\maketitle

\section{Introduction}

Quantum superposition is one of the most strange and intriguing concepts 
in quantum mechanics and a useful resource for quantum information processing. 
Superpositions of macroscopically distinct states are often referred to as Schr\"odinger cat states, 
or cat states for short, named after Schr\"odinger's famous gedankenexperiment with a cat 
in a superposition of alive and dead states~\cite{Haroche2013a,Wineland2013a}. 
In quantum optics, superpositions of distinct coherent states are called cat states~\cite{Leonhardt}, 
because coherent states are often regarded as the ``most classical" states of light.

Such cat states have been generated experimentally by various approaches. 
In the optical regime, cat states of small size, which are sometimes called Schr\"odinger kittens, 
have been generated by subtracting one photon from squeezed states of light~\cite{Ourjoumtsev2006a,Wakui2007a}. 
Optical cat states of a little larger size have been generated by other methods~\cite{Ourjoumtsev2007a,Sychev2017a}. 
Note that these optical cat states are of \textit{traveling} fields and 
generated~\textit{probabilistically}.

In the microwave regime, 
cat states of larger size have been generated experimentally~\cite{Deleglise2008a,Vlastakis2013a,Leghtas2015a,Touzard2018a}. 
The generation using Rydberg atoms~\cite{Deleglise2008a} 
is heralded by measurement results of the atomic states,
where the parity, `even' or `odd', of the cat state
is determined randomly according to the measurement results. 
\textit{On-demand} generations of microwave cat states have been demonstrated using superconducting circuits 
by two different approaches, one of which is based on conditional operations
using a superconducting quantum bit (qubit)~\cite{Vlastakis2013a} and the other is 
based on two-photon driving and two-photon loss larger than one-photon loss~\cite{Leghtas2015a,Touzard2018a}. 
By extending the former approach to a two-cavity case, 
entangled coherent states in two cavities have been observed experimentally~\cite{Wang2016a}. 
Note that these microwave cat states are confined \textit{inside} cavities. 
Recently, 
the cat state generated inside a cavity
has been released by controlling the output coupling rate of the cavity 
using four-wave mixing in the qubit~\cite{Pfaff2017a}. 
To the best of our knowledge, 
only this experiment has demonstrated \textit{on-demand} generation of \textit{traveling} cat states.

In this paper, 
we propose a simple alternative method for on-demand generation of traveling cat states. 
Our method is based on a recent theoretical result that a Kerr-nonlinear parametric oscillator (KPO) 
can generate intracavity cat states deterministically via quantum adiabatic evolution~\cite{Goto2016a,Puri2017a}. 
(The KPO has recently attracted attention for its application to quantum computing~\cite{Goto2016a,Puri2017a,Goto2016b,Nigg2017a,Puri2017b}.) 
In the previous work, the KPO is assumed to be lossless in ideal cases, 
and therefore the cat states are confined inside the KPO. 
In the present work, 
we theoretically investigate a coupled system of a KPO and a one-dimensional system (output mode). 
It turns out that 
the cat states generated inside a KPO can be released into the output mode 
by controlling the parametric pump amplitude dynamically,
while the output coupling rate is constant.
Hence by using a KPO,
we can generate \textit{traveling} cat states \textit{on demand} 
without controlling the output coupling rate.

\section{Model}

\begin{figure}[b]
	\includegraphics[width=8cm]{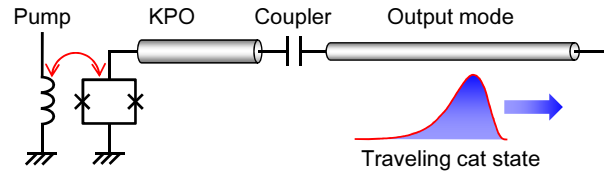}
	\caption{Coupled system of a KPO and an output mode.
	The KPO is implemented by a Josephson parametric oscillator (JPO)~\cite{Lin2014a}
	with a relatively large Kerr effect.
	The JPO is capacitively coupled to a transmission line for the output mode.
	Traveling cat state is generated by controlling the pump field dynamically.}
	\label{fig-system}
\end{figure}

The coupled system of a KPO and an output mode is depicted in Fig.~\ref{fig-system}, 
where a superconducting-circuit implementation is supposed~\cite{Nigg2017a,Puri2017b,Lin2014a}.
In a frame rotating at half the pump frequency, $\omega_p /2$, 
of the parametric pumping and in the rotating-wave approximation,
the system is modeled by the following Hamiltonian~\cite{Goto2016a,Puri2017a,Gardiner1985a,Milburn,Duan2003a,Goto2005a}
(we use the units $\hbar = v_p =1$, where $v_p$ is 
the phase velocity of the electromagnetic fields in the output mode):
\begin{align}
H(t)
&=
H_{\mathrm{KPO}}(t)+H_{\mathrm{out}}+H_{c},
\label{eq-H}
\\
H_{\mathrm{KPO}}(t)
&=
\frac{p(t)}{2} 
\left(
a^{\dagger 2} + a^2
\right)
-
\frac{K}{2} a^{\dagger 2} a^2
+
\Delta a^{\dagger} a,
\label{eq-HKPO}
\\
H_{\mathrm{out}}
&=
\int_{-\infty}^{\infty} \! \omega b^{\dagger} (\omega ) b(\omega ) \, d\omega,
\label{eq-Hout}
\\
H_c
&=
i 
\sqrt{\frac{\kappa_{\mathrm{ex}}}{2\pi}}
\int_{-\infty}^{\infty} \! 
\left[
b^{\dagger} (\omega ) a - a^{\dagger} b(\omega )
\right] d\omega,
\label{eq-Hc}
\end{align}
where $a^{\dagger}$ and $a$ are the creation and annihilation operators for the KPO, 
$p(t)$ is the time-dependent pump amplitude,
$K$ is the magnitude of the Kerr coefficient~\cite{comment-Kerr}, 
$\Delta=\omega_{\mathrm{KPO}}-\omega_p/2$  is the detuning frequency 
($\omega_{\mathrm{KPO}}$ is the one-photon resonance frequency of the KPO), 
$b^{\dagger} (\omega)$ and $b(\omega)$ are 
the creation and annihilation operators for photons of frequency (or wave number)
$\omega_p/2+\omega$ 
in the output mode, and 
$\kappa_{\mathrm{ex}}$ is the energy decay rate of the KPO due to its coupling to the output mode. 
Here we assume no internal loss of the KPO, 
which is discussed later.
Hereafter, we consider the resonance case ($\Delta=0$).

If the KPO is a closed system (${\kappa_{\mathrm{ex}}=0}$),
a cat state ${|\alpha_0 \rangle + |{-\alpha_0} \rangle}$
can be generated from the vacuum state $|0\rangle$ 
via quantum adiabatic evolution 
by gradually increasing $p(t)$ from zero to ${p_0=K \alpha_0^2}$~\cite{Goto2016a,Puri2017a}.
When the KPO is coupled to the output mode,
the photons inside the KPO will leak to the output mode.
As a result, the entanglement between the KPO and the output mode arises 
during the generation.
Moreover, the decay of the KPO due to the leak
may degrade the adiabatic cat-state generation.
Thus, it is not obvious whether or not we can generate a traveling cat state 
using the KPO.

\section{Proposed method}

Our idea is based on the fact that
any quantum state inside a \textit{linear} cavity results in a traveling pulse in the same quantum state
through the leak to the output mode,
where the pulse shape is exponential corresponding to the exponential decay~\cite{Eichler2011a}.
This property of linear cavities removes the concern with the entanglement.
The issue with the decay can also be solved
by generating a cat state
faster than the decay,
which is possible if $K$ is much larger than $\kappa_{\mathrm{ex}}$.
The remaining problem is that 
the KPO has a large Kerr effect, namely, it is not a linear cavity.

Our solution is to switch off the parametric pumping as 
${p(t) \propto \exp (-\kappa_{\mathrm{ex}} t)}$
after the cat-state preparation.
Then the Kerr term and the pumping term are cancelled out each other,
and hence the KPO can be regarded as a linear cavity.
This is confirmed as follows.
Suppose that at time $t_0$, the KPO is prepared in a cat state
${|\alpha_0 \rangle + |{-\alpha_0} \rangle}$,
where $\displaystyle {\alpha_0 = \sqrt{p_0/K}}$ and ${p_0 = p(t_0)}$.
Since $H_{\mathrm{KPO}}(t)$ in Eq.~(\ref{eq-HKPO}) is rewritten as 
$\displaystyle {H_{\mathrm{KPO}}(t) = -\frac{K}{2} 
\! \left( a^{\dagger 2} - \frac{p(t)}{K} \right)
\! \left( a^{2} - \frac{p(t)}{K} \right)}$ by dropping a c-number term,
${H_{\mathrm{KPO}}(t_0) |{\pm \alpha_0} \rangle \propto \alpha_0^2 - p_0/K = 0}$,
where ${a |{\pm \alpha_0} \rangle = \pm \alpha_0 |{\pm \alpha_0} \rangle}$~\cite{Leonhardt}.
Thus at $t_0$, the amplitude starts decreasing 
as $\displaystyle {\pm \alpha (t) = \pm \alpha_0 e^{-\kappa_{\tiny{\mbox{ex}}} (t-t_0)/2}}$
because of the external coupling.
If we set the pump amplitude as $p(t) = p_0 e^{-\kappa_{\tiny{\mbox{ex}}} (t-t_0)}$,
$H_{\mathrm{KPO}}(t) |{\pm \alpha (t)} \rangle \propto \alpha (t)^2 - p(t)/K = 0$,
that is, the Kerr term and the pumping term are cancelled out at any time $t \ge t_0$.
Thus, the KPO behaves like a linear cavity during the release of the cat state,
and therefore the cat state prepared inside the KPO is faithfully released as a traveling cat state.

\section{Numerical simulation}

To examine the above method quantitatively,
we numerically solve the Schr\"odinger equation with the Hamiltonian in Eq.~(\ref{eq-H}) and 
evaluate the fidelity between the output pulse and an ideal cat state.
An approach to the numerical simulation is based on the discretization of frequency $\omega$
(or wave number $\omega /v_p$)~\cite{Duan2003a,Goto2005a}.
In the present case, however,
the photon number in the one-dimensional system is larger than one,
and consequently such discretization results in a complicated system of differential equations.
In this work, we instead discretize the position,
which results in simpler equations as shown below.

We first introduce the annihilation operator with respect to the position $z$:
\begin{align}
\tilde{b}(z)
&=
\frac{1}{\sqrt{2\pi}}
\int_{-\infty}^{\infty} \!
b(\omega) e^{i \omega z} \, d\omega.
\end{align}
Then, we move to the interaction picture with the unitary operator $U(t)=e^{-iH_{\mathrm{out}}t}$ as follows:
\begin{align}
\tilde{b}_I(z,t)
&=
U^{\dagger}(t) \tilde{b}(z) U(t)
=
\tilde{b}(z-t),
\nonumber
\\
H_I(t)
&= 
U^{\dagger}(t) \left[ H_{\mathrm{KPO}}(t)
+ H_c \right] U(t)
\nonumber
\\
&=
H_{\mathrm{KPO}}(t)
+
i \sqrt{\kappa_{\mathrm{ex}}}
\left[
\tilde{b}^{\dagger}(-t) a
-
a^{\dagger} \tilde{b}(-t)
\right].
\nonumber
\end{align}
A pulse-mode operator for the interval $[0,T]$ is defined as 
\begin{align}
b_p = \int_0^{T} \! f_p(z) \tilde{b}_I(z,T) \, dz
=
\int_0^{T} \! f_p(T - z) \tilde{b}(-z) \, dz,
\nonumber
\end{align}
where $f_p(z)$ is the normalized envelope function for the output pulse satisfying
$\displaystyle \int_{0}^T \! |f_p(z)|^2 dz =1$.
In the present work,
we define $f_p(z)$ as ${f_p(z) \propto \sqrt{\langle \tilde{b}^{\dagger}_I(z,T) \tilde{b}_I(z,T) \rangle}}$,
where $\langle \tilde{b}^{\dagger}_I(z,T) \tilde{b}_I(z,T) \rangle$ is 
the spatial distribution of photons in the output mode at the final time $T$.
Note that in experiments, we can find such $f_p(z)$ from the measurement of the output power from the KPO.

Next, we divide the interval $[0,T]$ into $J$ small intervals $[z_{j-1}, z_{j}]$ 
($j=1, 2, \ldots, J$), where 
the intervals $\Delta z_j = z_{j}-z_{j-1}$ are set to small values~\cite{supplement}.
Then, $\tilde{b}(z)$ and $f_p(z)$ are discretized as follows
($z\in [z_{j-1}, z_{j}]$):
\begin{align}
\tilde{b}(-z) \sqrt{\Delta z_j} 
\to \tilde{b}_j,~
f_p(T-z) \sqrt{\Delta z_j} 
\to f_j.
\end{align}
Then, the commutation relation $[\tilde{b}(z),\tilde{b}^{\dagger}(z')]=\delta (z-z')$ 
becomes $[\tilde{b}_j, \tilde{b}^{\dagger}_l]=\delta_{j,l}$ and
the normalization condition $\displaystyle \int_{0}^T \! |f_p(z)|^2 dz =1$
becomes $\displaystyle \sum_{j=1}^J |f_j|^2 =1$.
By transforming the integration with respect to $z$ to the summation with respect to $j$,
we obtain
\begin{align}
b_p = \sum_{j=1}^{J} f_j \tilde{b}_j,~
f_j = 
\sqrt{\frac{\langle \tilde{b}_j^{\dagger} \tilde{b}_j \rangle}
{\sum_{l=1}^{J} \langle \tilde{b}_l^{\dagger} \tilde{b}_l \rangle}}.
\label{eq-bp}
\end{align}

The Hamiltonian at time $t \in [z_{j-1}, z_j]$
is given by
\begin{align}
H_I(t) 
&= 
H_{\mathrm{KPO}}(t)
+
i \sqrt{\kappa_{\mathrm{ex}}}
\left[
\tilde{b}_j^{\dagger} a
-
a^{\dagger} \tilde{b}_j
\right].
\label{eq-HI}
\end{align}
We numerically solve the Schr\"odinger equation with the Hamiltonian in Eq.~(\ref{eq-HI})~\cite{supplement}.
Since the Hamiltonian includes only one of $\{ \tilde{b}_j \}$,
the corresponding Schr\"odinger equation is simple.
In the present work,
we investigate the cases where $\kappa_{\mathrm{ex}}=0.2K$.

As explained above,
$p(t)$ should satisfy the following two conditions:
$p(t)$ is increased fast enough to adiabatically generate a cat state inside the KPO
before the decay spoils it;
after that,
$p(t)$ is decreased as $\displaystyle {p(t) \propto \exp (-\kappa_{\mathrm{ex}} t)}$
so that the Kerr term and the pumping term are cancelled out.
To satisfy these conditions simultaneously,
we define $p(t)$ as the output of the fourth order low-pass filter (LPF)~\cite{supplement,Wiseman}
with the input ${p_{\mathrm{in}}(t)=K A_p \exp (-\kappa_{\mathrm{ex}} t)}$.
The dimensionless parameter $A_p$ is used for tuning the photon number of the traveling cat state.
We set the photon number to about 2,
which is large enough for the two coherent states being distinct
but small enough to solve the Schr\"odinger equation numerically.
The bandwidth, $B$, of the LPF~\cite{supplement,Wiseman} is set to $B=0.5K$
between $\kappa_{\mathrm{ex}}$ and $K$.
The final time $T$ is set such that
the final photon number, $n_{\mathrm{in}}$, in the KPO is less than $10^{-3}$
(see Table~\ref{table-result}).

We obtain the density operator, $\rho$, describing the quantum state of the output pulse
using the moments 
$M_{m,n}=\langle b_p^{\dagger m} b_p^n \rangle$
with $b_p$ in Eq.~(\ref{eq-bp}).
The density matrix with respect to the Fock states 
is given by the following formula~\cite{Eichler2012a}:
\begin{align}
\rho_{m,n}
= \frac{1}{\sqrt{m!n!}}
\sum_{l=0}^{\infty} \frac{(-1)^l}{l!} M_{n+l,m+l}.
\label{eq-rho}
\end{align}
Using the density matrix,
we calculate the corresponding Wigner function
$\displaystyle {W(\beta ) = 
\frac{2}{\pi} \mbox{Tr} \! \left[ D(-\beta) \rho D(\beta) P \right]}$~\cite{Leonhardt,Deleglise2008a,Goto2016a},
where $\displaystyle {D(\beta)=\exp \! \left( \beta b_p^{\dagger} - \beta^* b_p \right)}$ 
(the asterisk denotes complex conjugation) and 
$\displaystyle {P=\exp \! \left( i\pi b_p^{\dagger} b_p \right)}$.

The results of the numerical simulation are shown in Fig.~\ref{fig-result}(a) and 
the first row of Table~\ref{table-result}.
Table~\ref{table-result} also provides the setting of the simulation.
Figure \ref{fig-result}(a) shows that 
the photon number in the KPO varies in a similar manner to the pump amplitude,
as expected.
The Wigner function for the output pulse in Fig.~\ref{fig-result}(a)
clearly shows the interference fringe, which is the evidence 
for the quantum superposition of the two coherent states,
that is, the output pulse is in a cat state.
(The tilt of the Wigner function, which corresponds to $\theta_{\mathrm{cat}}$ in Table~\ref{table-result},
is due to the residual Kerr effect.)
As shown in Table~\ref{table-result},
the maximum fidelity between this output state and an ideal cat state is 0.962.
Thus, the present method works successfully as expected.

\begin{widetext}

\begin{figure}[htbp]
	\includegraphics[width=15cm]{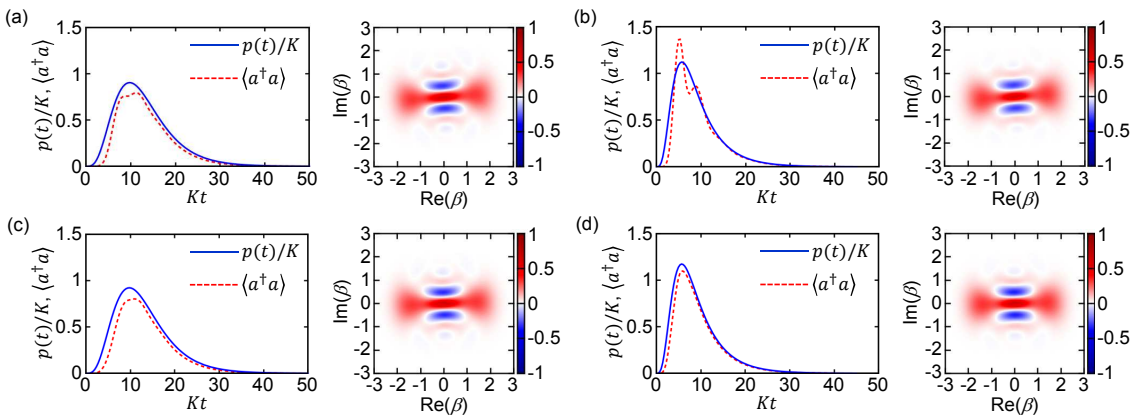}
	\caption{Simulation results of traveling cat-state generation using a KPO.
	Left: Time evolutions of the pump amplitude $p(t)$ and 
	the expectation value of the photon number, $\langle a^{\dagger} a \rangle$, in the KPO.
	Right: Wigner function, $W(\beta )$, of the output pulse.
	See Table~\ref{table-result} for the settings of (a)--(d).}
	\label{fig-result}
\end{figure}

\begin{table}[htb]
	\caption{Results and setting of numerical simulations.
	Fidelity: maximum fidelity between the output state and the ideal cat state 
	with two parameters:
	$\displaystyle (|\beta_{\mathrm{cat}} e^{i\theta_{\tiny{\mbox{cat}}}} \rangle
	+|{-\beta_{\mathrm{cat}} e^{i\theta_{\tiny{\mbox{cat}}}}} \rangle )
	/\sqrt{2(1+e^{-2\beta_{\tiny{\mbox{cat}}}^2})}$, where
	$\beta_{\mathrm{cat}}$ and $\theta_{\mathrm{cat}}$ are
	the magnitude and phase of the cat-state amplitude, respectively.
	$\beta_{\mathrm{cat}}^2$ and $\theta_{\mathrm{cat}}/\pi$:
	values maximizing the fidelity.
	$\beta_{\mathrm{cat}}^2$ is close to 2, which means
	that the photon number of the output pulse is about 2.
	$n_{\mathrm{in}}$: average photon number in the KPO at the final time $T$.
	$I_t$: time integral of the average photon number in the KPO, that is, $I_t=\int_0^T \! \langle a^{\dagger} a \rangle \, dt$.
	In all the cases, ${I_t \approx 10 K^{-1}=2\kappa_{\mathrm{ex}}^{-1}}$,
	because ${\kappa_{\mathrm{ex}} I_t}$ corresponds to the photon number of the output pulse.}
	\begin{tabular}{l|ccccc|cccccc} \hline \hline
	& ~Fidelity~ & $~~\beta_{\mathrm{cat}}^2~~$ 
	& $~~\theta_{\mathrm{cat}}/\pi~~$ & $n_{\mathrm{in}}$ & $~~K I_t~~$
	& ~Shortcut~ & $~~\kappa_{\mathrm{ex}}/K~~$ 
	& $~~B/K~~$ & $~~A_p~~$ & $~~KT~~$ & $~~J~~$ 
	\\ \hline 
	Fig.~\ref{fig-result}(a) & 0.962 & 2.01 & 0.03 & $6.2\times 10^{-4}$ & 9.63 & Unused & 0.2 & 0.5 & 2.45 & 50 & 80
	\\
	Fig.~\ref{fig-result}(b) & 0.930 & 1.96 & 0.02 & $6.1\times 10^{-4}$ & 9.64 & Unused & 0.2 & 1.0 & 2.15 & 45 & 80
	\\
	Fig.~\ref{fig-result}(c) & 0.983 & 2.03 & 0.02 & $6.4\times 10^{-4}$ & 9.63 & Used & 0.2 & 0.5 & 2.50 & 50 & 80
	\\
	Fig.~\ref{fig-result}(d) & 0.993 & 2.02 & 0.01 & $6.4\times 10^{-4}$ & 9.60 & Used & 0.2 & 1.0 & 2.25 & 45 & 80
	\\ \hline \hline
	\end{tabular}
	\label{table-result}
\end{table}

\end{widetext}

\section{Improvement by shortcut to adiabaticity}

The imperfection of the generated traveling cat state
may be partially due to the leak during the initial cat-state preparation.
We can speed up the preparation by setting the LPF bandwidth $B$ to a larger value,
e.g., $B=K$. Then, however, 
nonadiabatic effects degrade the cat state.
The simulation results for $B=K$
are shown in Fig.~\ref{fig-result}(b) and the second row of Table~\ref{table-result}.
The oscillation of $\langle a^{\dagger} a \rangle$ in Fig.~\ref{fig-result}(b)
is due to the nonadiabatic effects.
Consequently, the maximum fidelity between the output state and an ideal cat state
decreases to 0.930 (see Table~\ref{table-result}).

To mitigate the nonadiabatic effects,
we can use the technique called \textit{shortcut to adiabaticity}~\cite{Puri2017a,Demirplak2003a,Campo2013a}.
To maintain quantum adiabatic evolution,
the shortcut-to-adiabaticity technique introduces the following counterdiabatic Hamiltonian~\cite{Demirplak2003a,Campo2013a}:
\begin{align}
H_{\mathrm{counter}} = 
i \sum_n |\dot{\phi}_n \rangle \langle \phi_n|,
\end{align}
where $|\phi_n \rangle$ is the $n$-th instantaneous eigenstate of the slowly varying Hamiltonian 
and the dot denotes the time derivative.

In the present case,
the counterdiabatic Hamiltonian is approximately given by~\cite{supplement,comment-shortcut}
\begin{align}
H_{\mathrm{counter}} (t)
&= 
i \frac{p'(t)}{2} \left( a^{\dagger 2} - a^2 \right),
\label{eq-counter}
\\
p' (t)
&= 
\frac{\dot{p}(t)}{p(t)} \tanh \! \frac{p(t)}{K}.
\label{eq-pi}
\end{align}
The physical meaning of the counterdiabatic Hamiltonian
is to add the imaginary pump amplitude $p'(t)$ to the real one $p(t)$.
This is experimentally possible by controlling the phase of the pump field.

The simulation results with the shortcut-to-adiabaticity technique are shown in Figs.~\ref{fig-result}(c) and \ref{fig-result}(d)
and the third and fourth rows of Table~\ref{table-result}.
As shown in Fig.~\ref{fig-result}(d), the oscillation of $\langle a^{\dagger} a \rangle$ does not occur
even when $B=K$, unlike Fig.~\ref{fig-result}(b).
This demonstrates that the shortcut-to-adiabaticity technique works successfully.
The fidelity is improved for both $B=0.5K$ and $K$.
Contrary to the results without the shortcut-to-adiabaticity technique,
the fidelity is higher for larger $B$, and 
the corresponding infidelity is lower than 1\% when $B=K$
(See Table~\ref{table-result}).
Thus, the shortcut-to-adiabaticity technique can significantly improve the quality of the traveling cat state.

\section{Internal loss}

So far, internal loss of the KPO has not been taken into account.
However, any actual devices have internal loss,
and it degrades the coherence of the cat states.
Here we briefly examine the effect of the internal loss
in the superconducting-circuit implementation of the KPO~\cite{Nigg2017a,Puri2017b,Lin2014a}.

Assuming that $K/(2\pi)=10$~MHz and 
$\omega_{\mathrm{KPO}}/(2\pi)=10$~GHz as typical values,
$\kappa_{\mathrm{ex}}=0.2K$ in the present simulations
corresponds to the external quality factor,
$Q_{\mathrm{ex}}=\omega_{\mathrm{KPO}}/\kappa_{\mathrm{ex}}$, 
of $5\times 10^3$.
On the other hand,
the probability of losing a photon inside the KPO,
which gives an upper bound on the infidelity due to the internal loss, 
is approximately given by
$\kappa_{\mathrm{in}} I_t$,
where $\kappa_\mathrm{in}$
is the internal-loss rate and 
${I_t=\int_0^T \! \langle a^{\dagger} a \rangle \, dt}$.
In Table~\ref{table-result},
${I_t \approx 10 K^{-1}=2\kappa_{\mathrm{ex}}^{-1}}$ in all the cases.
Thus, in order to have an intra-KPO photon-loss probability below, e.g., 10\%, 
the required condition is
$\kappa_{\mathrm{in}} I_t \approx 10 \kappa_{\mathrm{in}}/K \leq 0.1$,
which is equivalent to the internal quality factor of $Q_{\mathrm{in}} \geq 10^5$.
These values of $Q_{\mathrm{ex}}$ and $Q_{\mathrm{in}}$
seem feasible with current technologies.

\section{Conclusion}

We have shown that the cat states deterministically generated inside a KPO
via quantum adiabatic evolution
can be released into an output mode by controlling the pump amplitude properly.
Thus, on-demand generation of traveling cat states can be realized using a KPO.
We have further shown that 
a shortcut-to-adiabaticity technique, where the phase of the pump field is controlled dynamically in time,
can improve the quality of the traveling cat state significantly.
The traveling cat states generated by a KPO can be directly observed 
by, e.g., homodyne or heterodyne detection of the output field.
Thus, this can be used for experimentally demonstrating 
the ability of a KPO to generate cat states deterministically.

\section*{Acknowledgments}
HG thanks Kazuki Koshino for his suggestion.
This work was supported by JST ERATO (Grant No. JPMJER1601).

\begin{widetext}

\begin{appendix}

\section{Numerical simulation method}

In the simulation presented in the main text,
we truncate the photon number in the output mode at 6,
which is sufficiently large to express a cat state with average photon number of 2.
Then, the state vector $|\psi \rangle$ describing the coupled system is represented as follows:
\begin{align}
|\psi \rangle
&=
\sum_{n=0}^{N_0} \psi_0 (n) |n\rangle |0\rangle
+
\sum_{n=0}^{N_1} \sum_{j_1=1}^{J} \psi_1 (n,j_1) |n\rangle |j_1\rangle
+
\sum_{n=0}^{N_2} \sum_{j_1=1}^{J} \sum_{j_2=1}^{j_1} \psi_2 (n,j_1,j_2) |n\rangle |j_1,j_2\rangle
\nonumber
\\
&+
\sum_{n=0}^{N_3} \sum_{j_1=1}^{J} \sum_{j_2=1}^{j_1} \sum_{j_3=1}^{j_2} 
\psi_3 (n,j_1,j_2,j_3) |n\rangle |j_1,j_2,j_3\rangle
+
\sum_{n=0}^{N_4} \sum_{j_1=1}^{J} \sum_{j_2=1}^{j_1} \sum_{j_3=1}^{j_2} \sum_{j_4=1}^{j_3} 
\psi_4 (n,j_1,j_2,j_3,j_4) |n\rangle |j_1,j_2,j_3,j_4\rangle
\nonumber
\\
&+
\sum_{n=0}^{N_5} \sum_{j_1=1}^{J} \sum_{j_2=1}^{j_1} \sum_{j_3=1}^{j_2} \sum_{j_4=1}^{j_3} \sum_{j_5=1}^{j_4} 
\psi_5 (n,j_1,j_2,j_3,j_4,j_5) |n\rangle |j_1,j_2,j_3,j_4,j_5\rangle
\nonumber
\\
&+
\sum_{n=0}^{N_6} \sum_{j_1=1}^{J} \sum_{j_2=1}^{j_1} \sum_{j_3=1}^{j_2} \sum_{j_4=1}^{j_3} \sum_{j_5=1}^{j_4} \sum_{j_6=1}^{j_5} 
\psi_6 (n,j_1,j_2,j_3,j_4,j_5,j_6) |n\rangle |j_1,j_2,j_3,j_4,j_5,j_6\rangle,
\label{eq-psi}
\end{align}
where the first and second ket vectors represent the Fock states of the KPO and the output mode, respectively,
and $N_l$ ($l=0, 1, \ldots , 6$) is the number at which the photon number in the KPO is truncated when 
the photon number in the output mode is $l$.
In the present simulations,
we set $N_0=N_1=N_2=6$, $N_3=5$, $N_4=4$, $N_5=3$, and $N_6=2$. 
The second ket vector is defined with the creation operators, e.g., as follows:
\begin{align}
|j_1,j_2,j_3\rangle
=\mathcal{N}(j_1,j_2,j_3) \tilde{b}^{\dagger}_{j_1} \tilde{b}^{\dagger}_{j_2} \tilde{b}^{\dagger}_{j_3} |0\rangle,
\end{align}
where the normalization factor $\mathcal{N}$ is defined as
\begin{align}
\mathcal{N}(j_1,j_2,j_3)
=
\left\{
\begin{matrix}
1 & \cdots & j_1>j_2>j_3, \\
1/\sqrt{2!} & \cdots & j_1=j_2>j_3, \\
1/\sqrt{2!} & \cdots & j_1>j_2=j_3, \\
1/\sqrt{3!} & \cdots & j_1=j_2=j_3. \\
\end{matrix}
\right.
\end{align}
Using this representation,
we numerically solved the Schr\"odinger equation with the Hamiltonian in Eq.~(8) in the main text.
Since the Hamiltonian includes only one of $\{ \tilde{b}_j \}$,
the corresponding Schr\"odinger equation is simple.
Moreover, at time $t \in [z_{j-1}, z_j]$,
$|\psi \rangle$ includes only the output-mode photons satisfying
$j_1 \le j$ in Eq.~(\ref{eq-psi}).
This enables fast implementation of the simulation.

Since the photon number in the KPO is large (small) in the first (second) half of the whole process,
we set correspondingly the intervals $\Delta z_j$ to small (large) values.
More concretely,
we set $\Delta z_j=(T/2)/(4J/5)$ for $j=1, 2, \cdots, 4J/5$ and
$\Delta z_j=(T/2)/(J/5)$ for $j=4J/5+1, 4J/5+2, \cdots, J$.
(So we set $J$ to multiples of 5.)
We use the fourth-order Runge-Kutta method for numerically solving the Schr\"odinger equation,
where the time steps are set to about $0.1K^{-1}$.
These time steps are defined by dividing $\Delta z_j$ by appropriate integers.

As mentioned in the main text,
We set $J=80$ in the present simulations.
The following results show that this value of $J$ is sufficiently large.
Figure \ref{fig-J} shows the $J$ dependence of the final photon number, 
$n_{\mathrm{out}}=\sum_{j=1}^J \langle \tilde{b}_j^{\dagger} \tilde{b}_j \rangle$, 
in the output mode.
These data, the circles in Fig.~\ref{fig-J}, 
are well fitted with $n_0 - b/J$, the solid lines in Fig.~\ref{fig-J}, 
where $n_0$ and $b$ are the fitting parameters.
This is natural because the position discretization is the first-order approximation
with respect to $J^{-1}$.
The dashed lines in Fig.~\ref{fig-J} show $n_{\mathrm{out}}=n_0$,
which are the estimated values of $n_{\mathrm{out}}$ in the limit $J \to \infty$.
From the small discrepancies between the data and $n_0$, indicated by arrows in Fig.~\ref{fig-J},
the numerical errors due to finite $J$ are estimated to be less than 1\%.
This indicates that $J$ is sufficiently large.

\begin{figure}[t]
	\includegraphics[width=12cm]{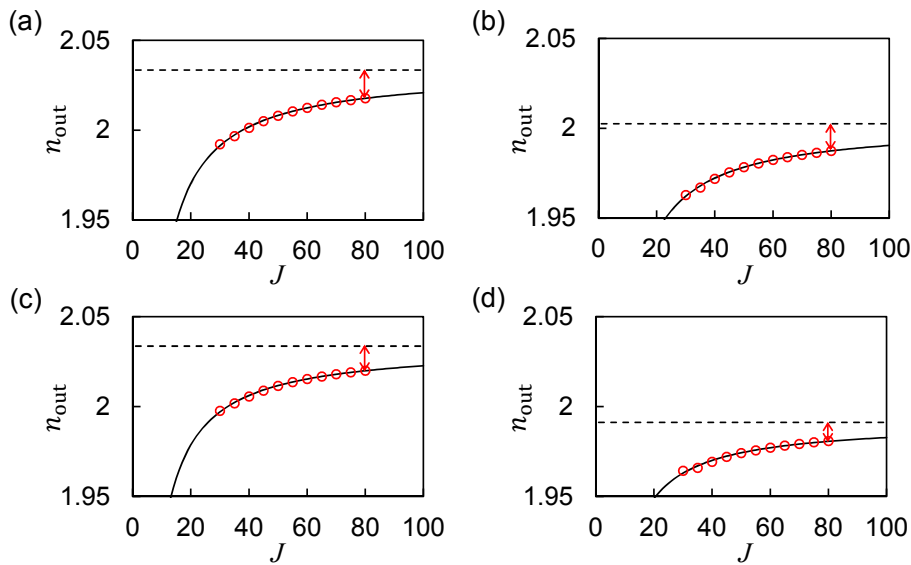}
	\caption{$J$ dependence of the final photon number, 
	$n_{\mathrm{out}}=\sum_{j=1}^J \langle \tilde{b}_j^{\dagger} \tilde{b}_j \rangle$,
	in the output mode.
	(a)--(d) correspond to Figs.~2(a)--2(d), respectively, in the main text.
	Circles represent simulation results.
	Solid lines are fitted curves with $n_{\mathrm{out}}(J) = n_0 - b/J$,
	where $n_0$ and $b$ are the fitting parameters.
	Horizontal dashed lines represent $n_{\mathrm{out}}(J) = n_0$.
	Arrows indicate the discrepancies between the data and $n_0$.}
	\label{fig-J}
\end{figure}

\section{Pulse shape control using low-pass filters}

In the simulation presented in the main text, 
we define the pulse shape of the pump amplitude $p(t)$ 
as the output of the fourth-order low-pass filter (LPF)
[29]
with the input ${p_{\mathrm{in}}(t)=K A_p \exp (-\kappa_{\scriptsize{\mbox{ex}}} t)}$.
The input-output relation of a LPF is given by 
$p_{\mathrm{out}}(t)
=\int_0^t \! B e^{-B(t-s)} p_{\mathrm{in}}(s) \, ds$
[29],
where $B$ is the bandwidth of the LPF and
$p_{\mathrm{in}}(s)=0$ (${s<0}$) is assumed.
Note that $\dot{p}_{\mathrm{out}}(t)
=-B [p_{\mathrm{out}}(t) - p_{\mathrm{in}}(t)]$,
where the dot denotes the time derivative.
Thus, we can calculate the output of the LPF by numerically solving this differential equation.
The $n$-th order LPF is defined as the output of the LPF
the input of which is the output of the $(n-1)$-th order LPF.

\section{Shortcut to adiabaticity for KPO}
\label{sec-shortcut}

Here we derive the approximate counterdiabatic Hamiltonian given by Eqs.~(11) and (12) in the main text
and provide numerical evidence for its validity.

First, using the completeness relation 
$\displaystyle \sum_n |\phi_n \rangle \langle \phi_n |=I$
($I$ is the identity operator),
the counterdiabatic Hamiltonian in Eq.~(10) in the main text is rewritten as follows:
\begin{align}
H_{\mathrm{counter}} = 
\frac{i}{2} \sum_n \left(
|\dot{\phi}_n \rangle \langle \phi_n| - |\phi_n \rangle \langle \dot{\phi}_n|
\right).
\label{eq-Hcounter0}
\end{align}

Among $\{ |\phi_n \rangle \}$,
we are interested only in the following even cat state:
\begin{align}
|{C_+ (p(t))} \rangle = 
\frac{\left| {\sqrt{p(t)/K}} \right\rangle + \left| {-\sqrt{p(t)/K}} \right\rangle}{\sqrt{2(1+e^{-2p(t)/K})}}.
\end{align}
Note that 
$\displaystyle H_{\scriptsize{\mbox{KPO}}}(t) |C_+ (p(t)) \rangle = \frac{p(t)^2}{2K} |C_+ (p(t)) \rangle$,
and therefore the even cat state is one of the energy eigenstates.
Disregarding the energy eigenstates other than $|C_+ \rangle$, 
the counterdiabatic Hamiltonian in Eq. (\ref{eq-Hcounter0}) is approximately given by
\begin{align}
H_{\mathrm{counter}} \approx 
\frac{i}{2}  \left(
|\dot{C}_+ \rangle \langle C_+| - |C_+ \rangle \langle \dot{C}_+|
\right).
\label{eq-Hcounter}
\end{align}

Using the odd cat state 
\begin{align}
|C_- (p(t)) \rangle = 
\frac{\left| {\sqrt{p(t)/K}} \right\rangle - \left| {-\sqrt{p(t)/K}} \right\rangle}{\sqrt{2(1-e^{-2p(t)/K})}},
\end{align}
$|\dot{C}_+ \rangle$ becomes
\begin{align}
|\dot{C}_+ \rangle = 
-\frac{\dot{p}}{2K} \tanh \! \frac{p}{K} \, |C_+ \rangle
+
\frac{\dot{p}}{2\sqrt{Kp}} \sqrt{\tanh \! \frac{p}{K}} \, a^{\dagger} |C_- \rangle.
\label{eq-dotCplus}
\end{align}

Substituting Eq.~(\ref{eq-dotCplus}) into Eq.~(\ref{eq-Hcounter}),
we obtain
\begin{align}
H_{\mathrm{counter}} 
\approx 
-\frac{i}{2}  \left(
\frac{\dot{p}}{2K} \tanh \! \frac{p}{K} \, 
|C_+ \rangle \langle C_+| 
+ 
\frac{\dot{p}}{2\sqrt{Kp}} \sqrt{\tanh \! \frac{p}{K}} \, 
|C_+ \rangle \langle C_-| a
\right)
+
\mbox{H.c.},
\label{eq-Hcounter2}
\end{align}
where H.c. denotes the Hermitian conjugate.

Using 
$\displaystyle a^2|C_+ \rangle = \frac{p}{K} |C_+ \rangle$
and 
$\displaystyle \left( |C_+ \rangle \langle C_-| a \right) |C_+ \rangle 
= \sqrt{\frac{p}{K} \tanh \! \frac{p}{K}} \, |C_+ \rangle
= \sqrt{\frac{K}{p} \tanh \! \frac{p}{K}} \, a^2 |C_+ \rangle$,
$H_{\mathrm{counter}}$ acting on $|C_+ \rangle$ 
is approximated as follows:
\begin{align}
H_{\mathrm{counter}} 
\approx 
-\frac{i}{2}  \left(
\frac{\dot{p}}{2p} \tanh \! \frac{p}{K} \, a^2
+ 
\frac{\dot{p}}{2\sqrt{Kp}} \sqrt{\tanh \! \frac{p}{K}} \, 
\sqrt{\frac{K}{p} \tanh \! \frac{p}{K}} \, a^2
\right)
+
\mbox{H.c.}
=
-\frac{i}{2} \frac{\dot{p}}{p} \tanh \! \frac{p}{K} \, a^2
+
\mbox{H.c.}
\label{eq-Hcounter3}
\end{align}
Thus, we obtain Eqs.~(11) and (12) in the main text.

To confirm the validity of the above derivation,
we performed numerical simulations of a KPO without the output coupling,
where the pump amplitude $p(t)$ is increased linearly from 0 to $2K$ at time $10K^{-1}$.
The results are summarized in Fig.~\ref{fig-shortcut}
together with the results in the cases without shortcut-to-adiabaticity technique
and with the shortcut-to-adiabaticity technique proposed in Ref.~15.
The technique proposed in Ref.~15 also uses 
an imaginary pump amplitude $p'(t)$,
but it is defined as
\begin{align}
p'(t)
=\frac{2\dot{\alpha}_0 \sqrt{1-2e^{-2|\alpha_0|^2}}}{1+2\alpha_0}
=
\frac{\dot{p} \sqrt{1-2e^{-2p/K}}}{\sqrt{Kp}+2p},
\label{eq-pi-Puri}
\end{align}
where $\alpha_0 = \sqrt{p/K}$ is used.
While the results with the technique in Ref.~15 exhibit oscillations,
the results with our technique change monotonically and the final fidelity is almost perfect.
These results clearly show the usefulness of our shortcut-to-adiabaticity technique for a KPO.

\begin{figure}[t]
	\includegraphics[width=12cm]{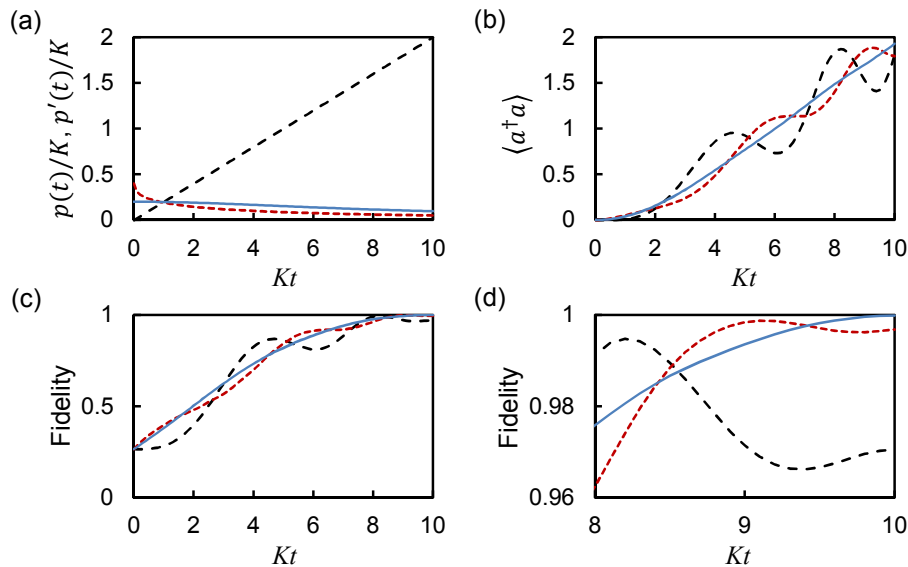}
	\caption{Simulation results for shortcut to adiabaticity.
	(a) Pump amplitudes, $p(t)$ (black long-dashed line), $p'(t)$ in our method (cyan solid line), and $p'(t)$ in Ref.~15 (red short-dashed line).
	(b) Average photon number in KPO, $\langle a^{\dagger} a \rangle$. 
	(c) Fidelity between the final state and the ideal even cat state with amplitude of $\sqrt{2}$.
	(d) Magnification of (c) around the final time.
	In (b)--(d), cyan solid, red short-dashed, and black long-dashed lines correspond to
	our method, the method proposed in Ref.~15, and the case of no $p'(t)$
	(without shortcut-to-adiabaticity technique), respectively.}
	\label{fig-shortcut}
\end{figure}

\end{appendix}

\end{widetext}


\begin{thebibliography}{99}
\bibitem{supplement}
See Appendix for 
the detailed methods and results of the present numerical simulations,
the pulse shape control of the pump amplitude using low-pass filters,
and the derivation of the counterdiabatic Hamiltonian 
for our shortcut-to-adiabaticity technique.
\end{thebibliography}
\end{document}